# The Frequency Dependence of Critical-velocity Behavior in Oscillatory Flow of Superfluid $^4$He Through a 2-μm by 2-μm Aperture in a Thin Foil

J. A. Flaten,[1] C. A. Lindensmith,[2] and W. Zimmermann, Jr.[3]

[1]*Department of Physics, Luther College, Decorah, IA 52101, USA*
[2]*Jet Propulsion Laboratory, Pasadena, CA 91109, USA*
[3]*Tate Laboratory of Physics, University of Minnesota, Minneapolis, MN 55455, USA*

**Abstract.** The critical-velocity behavior of oscillatory superfluid $^4$He flow through a 2-μm by 2-μm aperture in a 0.1-μm-thick foil has been studied from 0.36 K to 2.10 K at frequencies from less than 50 Hz up to above 1880 Hz. The pressure remained less than 0.5 bar. In early runs during which the frequency remained below 400 Hz, the critical velocity was a nearly-linearly decreasing function of increasing temperature throughout the region of temperature studied. In runs at the lowest frequencies, isolated $2\pi$ phase slips could be observed at the onset of dissipation. In runs with frequencies higher than 400 Hz, downward curvature was observed in the decrease of critical velocity with increasing temperature. In addition, above 500 Hz an alteration in supercritical behavior was seen at the lower temperatures, involving the appearance of large energy-loss events. These irregular events typically lasted a few tens of half-cycles of oscillation and could involve hundreds of times more energy loss than would have occurred in a single complete $2\pi$ phase slip at maximum flow. The temperatures at which this altered behavior was observed rose with frequency, from ~ 0.6 K and below, at 500 Hz, to ~ 1.0 K and below, at 1880 Hz.

## I. INTRODUCTION

A number of experiments carried out since the mid-1980's have shown that when superfluid $^4$He flows through a very small aperture in a thin membrane, its critical-velocity behavior is surprisingly regular, with several characteristic features. For apertures with lateral dimensions on the order of one micron or less in membranes one or two tenths of a micron in thickness, the critical velocity at the lowest temperatures is typically from one to ten meters per second. This velocity decreases with increasing temperature, often in a nearly linear fashion, over a wide range of temperatures. In addition, the critical velocity is highly reproducible within a given low-temperature run and is often fairly reproducible from run to run. Moreover, it has been shown that the dominant energy loss mechanism at the critical velocity is a series of independent $2\pi$ phase slips. It is believed that each such phase slip involves the nucleation and growth of a single quantum vortex to which the energy of potential flow is transferred. Multiple phase slips are sometimes observed.[1-12]

The behavior described above is in striking contrast to the critical-velocity behavior that is usually seen in more open or longer flow channels. In such channels, the critical velocity is usually relatively temperature-independent, except near the lambda temperature, where the critical velocity goes to zero. The critical velocity is typically less than 1 m/s and decreases with increasing lateral dimensions of the channel. Moreover, such critical-velocity behavior is often erratic and noisy and to some degree irreproducible.[2] Although some aspects of this behavior are accounted for qualitatively by a suggestion involving vortex production made by Feynman years ago[13] and by some results of the numerical simulation of vortex motion[14], a proper model for this behavior has never been developed, and the problem remains an interesting one.

Several years ago, one of us observed that when oscillatory critical-flow experiments were carried out with an aperture in the shape of a narrow slot 5 μm long and 0.3 μm wide in a foil 0.2 μm thick, regular critical behavior of the first type mentioned above was observed in a range of frequencies from 70 to 118 Hz as the temperature was lowered from 2.15 K down to 0.35 K. However, in a range of frequencies from 1818 to 1899 Hz, a different behavior set in as the temperature was decreased.[15] Below about 1.7 K, critical-velocity behavior at the higher frequency became noisy and metastable, and the critical velocity was noticeably less temperature-dependent than it was at the lower frequency. The noise appeared to be associated with successive erratic energy-loss events involving much more energy loss than could be accounted for by small numbers of independent single-vortex events.

The present experiments were undertaken to try to investigate this difference in behavior in a more complete and systematic fashion. The observations above suggested that we might be seeing the first steps of a transition from critical behavior of the first type described earlier to that of the second type. In any event, we believed that it would be worthwhile to investigate this new behavior further under the relatively well-controlled conditions of oscillatory flow experiments and to check that the original observations were not due to some experimental artifact associated with using two different modes of oscillation of the cell.

For this work we chose a single, approximately-square aperture with lateral dimensions of approximately 2 microns, large enough, we thought, for a transition of critical-velocity behavior with increasing frequency to be likely, in view of the results described above.[15] A series of experimental runs was made with this aperture at an increasing succession of frequencies, using only the lowest mode of oscillation of our experimental cell. In addition, a single final run was made with a configuration of the cell which allowed two modes of the cell to be used, as in the previous work, in order to check the dependence of critical-velocity behavior on frequency that was seen in the earlier runs of the present work.

These measurements were intended to be the first step in a series of measurements with a succession of apertures of increasing size, each at a number of frequencies, with a view to exploring and trying to understand more completely the transition in behavior noted above. A preliminary account of this work was given in Ref. 16. A more detailed but earlier account of much of this work can be found in Ref. 17.

## II. EXPERIMENT

The experimental cell, shown in Fig. 1, is a modification of one used in earlier work.[15,18,19] It consisted primarily of two chambers, upper and lower, connected by the small aperture under study. In some runs, a relatively large bypass channel connected the two chambers in parallel with the small aperture. In the primary mode of operation, the chambers were completely filled with superfluid $^4$He. The cell constituted a fluid-dynamical resonator in which oscillatory flow took place between the chambers through the opening or openings connecting them. This motion could be excited electrically by means of a piezoelectric driver attached to the flexible diaphragm that constituted the upper wall of the upper chamber. The motion could be sensed capacitively by means of the flexible diaphragm that constituted the lower wall of the lower chamber. The capacitor formed by this diaphragm and the fixed plate below it constituted the $C$ of an $LC$ back-diode rf oscillator operating at ~ 10 MHz. This oscillator was attached to the lower end of the cell, and its frequency-modulated output was detected outside the cryostat. The resonant oscillations of the fluid could also be excited by gently shaking the entire cell with a second piezoelectric element mounted on the outside of the cell or by an electromechanical vibrator mounted at the top of the dewar stand. When either of these alternative methods of excitation was used, the first piezoelectric driver mentioned above could serve as an ac detector of fluid motion. Although less easily calibrated from fundamental considerations, this latter detector sometimes yielded a better signal-to-noise ratio in the output signal than the capacitive detector.

The cell was suspended by springs, for vibration isolation, from the $^3$He pot of a conventional recirculating $^3$He cryostat. An additional "one-shot" $^3$He cooling stage was located on the cell itself for reaching the lowest temperatures. A low-temperature valve isolating the cell from its fill lines was located on the $^3$He pot. For further vibration isolation, the entire cryostat and dewar assembly was mounted on pneumatic vibration isolators. At times it also proved helpful to suppress vibration due to boiling in the liquid nitrogen bath by pumping on a copper tube fitted with a small inlet opening and immersed in the bath.

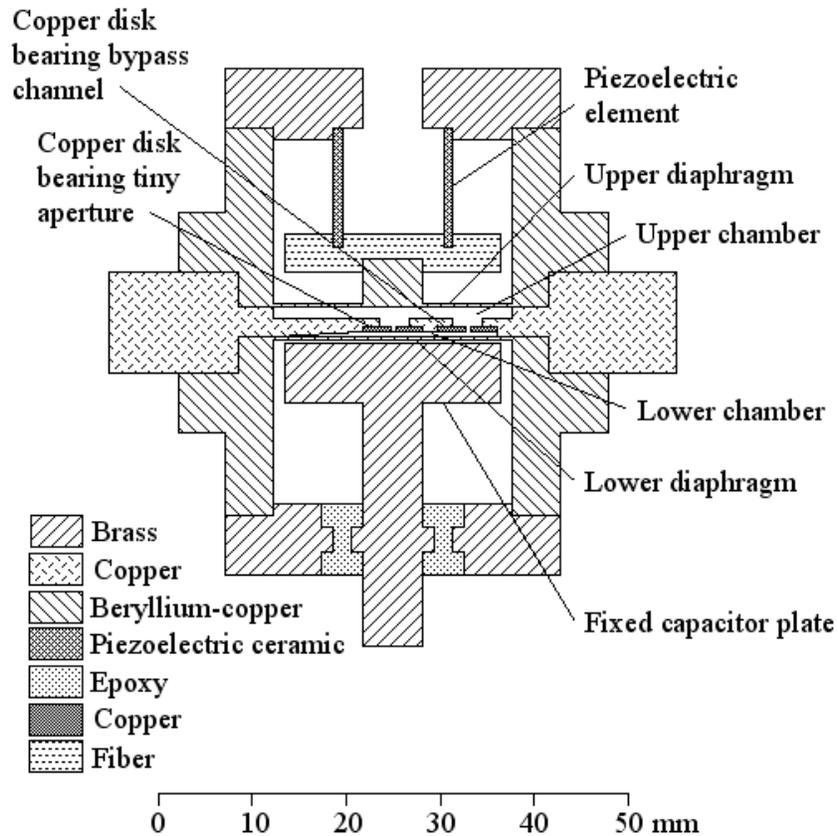

FIG. 1. Scale drawing of the experimental cell. The tiny aperture under study in the central copper disk is too small to be seen on this scale. A large bypass channel is shown in the right-hand copper disk. The spacing between the lower diaphragm and the fixed capacitor plate has been exaggerated.

The small aperture used in this work, shown in Fig. 2, was approximately square, measuring 2 μm on a side. It was etched in a 0.1-μm-thick titanium foil using electron-beam lithography of PMMA resist followed by plasma etching with $CCl_2F_2$. As seen in the figure, the edges of the aperture were quite ragged and could possibly have formed a multiply-connected flow channel.

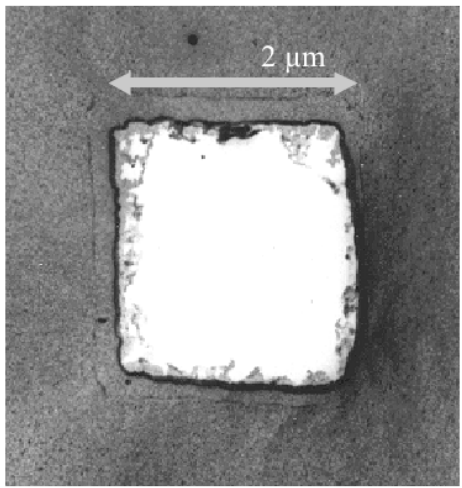

FIG. 2. Scanning electron micrograph of the tiny aperture studied in this set of experiments. The aperture has rough edges and measures approximately 2 μm by 2 μm. It is etched in a 0.1-μm-thick titanium foil.

The cell possessed several resonant modes of oscillation. The lowest mode, the one used for most of the critical-phase-difference determinations in this work, can be thought of as similar to a simple Helmholtz-like mode of a fluid oscillating through a constricted passage between two rigid-walled chambers. However, as discussed in Section III, the flexibility of the chamber walls in fact played an important role in our cell. At the lowest temperatures, with the small aperture alone connecting the chambers, the resonant frequency of this lowest mode was 73 Hz with the lower diaphragm shown in Fig. 1. Higher resonant frequencies for this mode were obtained by installing one of several different bypass channels in parallel with the small aperture. These channels included a 1.0-mm-long, 0.17-mm-i.d. capillary with a 0.10-mm-diameter wire insert and holes from 0.31 mm to 2.8 mm in diameter drilled in 0.5-mm-thick copper disks. These channels yielded low-temperature resonant frequencies from 186 Hz to 1448 Hz. The resonant frequencies of the lowest mode all decreased with increasing temperature, due to the decrease in superfluid density and the limited participation of the viscous normal fluid component in the oscillation. In all cases, the observed critical-flow phenomena occurred in the small aperture rather than in the bypass channel, as inferred from the observed supercritical shifts in frequency and changes in the shape of the resonant response.[18]

The cell also possessed higher-frequency resonant modes involving different phase relations between fluid motion and diaphragm motion. With the lower diaphragm shown in Fig. 1, the first of these modes occurred at frequencies near 8000 Hz. Neither this mode nor higher ones were used for critical-phase-difference determinations with this lower diaphragm, because the behavior of the cell at these frequencies remains open to some questions discussed later on. However, by installing a more-massive lower diaphragm that had been used in earlier work,[15,18,19] it was possible to obtain low-temperature resonance frequencies of 83 Hz and ~ 1880 Hz for the two lowest modes, with the small aperture alone. Critical-phase-difference data were collected in both of these modes in the last experimental run.

Additional frequencies of operation could be obtained by operating the cell in its lowest resonant mode with the upper chamber only partially filled. These frequencies lay below the corresponding completely-filled-cell frequencies. Although the fluid was largely decoupled from the upper diaphragm under these circumstances, the resonance could be excited by shaking the entire cell as described earlier.

As a first step in obtaining critical-response data at a particular temperature, the resonant frequency at that temperature was determined by measuring the response at some subcritical level of drive as the drive frequency was slowly swept through the resonance. Since the $Q$'s were typically > 10,000 at low temperature, the sweep times needed for true steady-state response could be quite long. We often found it expedient to sweep the frequency both up and down through the resonance and to identify the true resonance frequency as the point of symmetry between two mirror-image, skewed, non-steady-state resonance curves.[17]

Then, at the resonance frequency, it was usual to sweep the drive amplitude both upwards and downwards to locate the critical response level. This level characteristically appeared as an abrupt change in the slope of the response amplitude versus drive amplitude, a greater slope at subcritical amplitudes, a lesser slope at amplitudes above critical. Once the critical level was located, it was frequently useful to observe the response at a fixed, resonant drive frequency and a fixed, marginally-supercritical drive amplitude as a function of time. In this way, discrete $2\pi$ phase slips or larger collapses could be observed and studied.

### III. DETERMINATION OF CRITICAL PHASE DIFFERENCES AND VELOCITIES

In order to determine critical order-parameter phase-differences and critical velocities from critical response data, it was necessary to work out the relationship between the order-parameter phase-difference and the experimental response amplitude for subcritical velocities of flow. For this purpose we adopted a model for the cell that allows for elastic deformations of the upper and lower diaphragms and the accompanying inertias of these diaphragms, as well as for the compressibility and inertia of the fluid. The fluid is assumed to obey the standard two-fluid model equations of motion.[20] In our model for the cell, we assume that the pressures and temperatures are uniform within each chamber. Derivations of the results presented below, beginning with Eq. (6), are given in the Appendix.

In our model, the lower and upper diaphragms are assumed to obey, respectively, the equations of motion

$$\text{(lower diaphragm)} \quad m_1 \frac{d^2 x_1}{dt^2} = -k_1 x_1 - A_1 P_1 \quad \text{and} \quad (1)$$

$$\text{(upper diaphragm)} \quad m_2 \frac{d^2 x_2}{dt^2} = -k_2 x_2 + A_2 P_2 + \beta V. \quad (2)$$

Here the $x$'s are the displacements of the centers of the diaphragms from equilibrium in the upward direction, the $m$'s are the effective masses of the diaphragms, the $k$'s are the effective spring constants of the diaphragms, the $A$'s are the effective areas of the diaphragms, the $P$'s are the pressures in the chambers, $V$ is the drive voltage applied to the piezoelectric cylinder attached to the upper diaphragm, and $\beta$ is a constant drive coefficient. Throughout, the subscript 1 refers to the lower chamber and the subscript 2 to the upper chamber. The $m$'s and $k$'s are defined such that $m(dx/dt)^2/2$ equals the kinetic energy and $kx^2/2$ equals the elastic potential energy of the corresponding diaphragm. This model is similar to one adopted by Beecken and Zimmermann for the same cell, using the more-massive lower diaphragm, except that in their model, no provision was made for the compliance and inertia of the upper diaphragm.[18,19]

We begin by considering the low-temperature limit, in which the liquid consists entirely of the superfluid component and thermal effects can be neglected. Here we assume ideal, vortex-free flow of the superfluid. When two openings in parallel exist between chambers, we assume that any circulation present is constant. In this limit, the cell possesses three resonant modes of oscillation at angular frequencies that we designate, in order of increasing frequency, $\omega_-$, $\omega_+$, and $\omega_{++}$. These modes can be thought of as arising from the coupling of the ideal Helmholtz-like mode of oscillation of the liquid between chambers that would exist if the chamber walls were rigid, at angular frequency $\omega_H$, together with the separate empty-cell modes of vibration of the individual diaphragms, at angular frequencies $\omega_1$ and $\omega_2$. These uncoupled angular frequencies are given, respectively, by the expressions

$$\omega_H^2 = \frac{1}{\rho^2 \kappa L} \frac{\Omega_1 + \Omega_2}{\Omega_1 \Omega_2}, \qquad \omega_1^2 = \frac{k_1}{m_1}, \quad \text{and} \quad \omega_2^2 = \frac{k_2}{m_2}. \qquad (3)$$

Here $\rho$ is the density of the fluid, $\kappa$ is the compressibility of the fluid, $L$ is the overall fluid-dynamical "inductance" of the opening(s) between chambers, and $\Omega_1$ and $\Omega_2$ are the volumes of the lower and upper chambers, respectively. For future reference, $\nu_1 = \omega_1/2\pi$ equaled from 4200 to 4400 Hz for the lower diaphragm shown in Fig. 1 and 1143 Hz for the more-massive lower diaphragm used in the last experimental run. We believe that $\nu_2 = \omega_2/2\pi$ exceeded 10,000 Hz for this experimental cell.

The inductance $L$ is defined such that $LI^2/2$ equals the total kinetic energy of fluid flow through the opening(s) associated with the net mass current $I$ from chamber to chamber. For a single opening, $L$ can be expressed as

$$L = \frac{\ell_{eff}}{\rho S}, \qquad (4)$$

where $S$ is the minimal area spanning the opening and $\ell_{eff}$ is the effective fluid-dynamical length of the opening, defined for an ideal fluid by

$$\ell_{eff} = \frac{1}{v_{avg}} \int_1^2 \vec{v} \cdot d\vec{\ell}. \qquad (5)$$

Here $v_{avg}$ is the average of the normal component of the velocity over the minimal area spanning the opening, and the line integral is taken along any path that runs from a quiescent point in chamber 1 through the opening to a quiescent point in chamber 2. When two openings $a$ and $b$ are present in parallel, the overall $L$ is given by $L_a L_b / (L_a + L_b)$, just as for two electrical inductors in parallel.

Nearly all of our data were collected at the lowest resonant angular frequency $\omega_-$. In all of our experimental cases, $\omega_H$ was much less than $\omega_1$. Under the assumption that $\omega_H \ll \omega_1 \ll \omega_2$, when the cell is completely filled, $\omega_-$ is given in the low-temperature limit to very good approximation ($\leq 1\%$ error) by the formula

$$\omega_-^2 \cong \omega_{Heff}^2 \left( 1 - \frac{\alpha_1}{1+\alpha_1} \frac{\Omega_{2eff}}{\Omega_{1eff} + \Omega_{2eff}} \frac{\omega_{Heff}^2}{\omega_1^2} \right), \tag{6}$$

where we define

$$\omega_{Heff}^2 \equiv \frac{1}{\rho^2 \kappa L} \frac{\Omega_{1eff} + \Omega_{2eff}}{\Omega_{1eff} \Omega_{2eff}}. \tag{7}$$

Here we introduce effective volumes $\Omega_{1eff}$ and $\Omega_{2eff}$ that take into account the compliances of the diaphragms. These effective volumes are given by the expressions

$$\Omega_{1eff} \equiv \Omega_1 (1+\alpha_1) \quad \text{and} \quad \Omega_{2eff} \equiv \Omega_2 \left( 1 + \frac{\alpha_2}{1-(\omega^2/\omega_2^2)} \right), \tag{8}$$

where

$$\alpha_1 \equiv \frac{A_1^2}{\kappa k_1 \Omega_1} \quad \text{and} \quad \alpha_2 \equiv \frac{A_2^2}{\kappa k_2 \Omega_2}. \tag{9}$$

and where $\Omega_{2eff}$ is to be evaluated at $\omega_-$ in Eqs. (6) and (7). Thus Eq. (6) is an implicit expression for $\omega_-$. However, when $\omega_- \ll \omega_2$, the dependences of the right-hand sides of Eqs. (6) and (7) on $\omega_-$ can be neglected.

When the upper chamber of the cell is only partially filled, we have

$$\omega_-^2 = \omega_{Heff*}^2 \left( 1 - \frac{\alpha_1}{1+\alpha_1} \frac{\omega_{Heff*}^2}{\omega_1^2} \right), \tag{10}$$

where

$$\omega_{Heff*}^2 \equiv \frac{1}{\rho^2 \kappa L} \frac{1}{\Omega_{1eff}}, \tag{11}$$

reflecting the absence of pressure oscillation in the upper chamber and of any role for the upper diaphragm. As can be seen from Eqs. (6) and (10), when $\omega_{Heff}$ or $\omega_{Heff*} \ll \omega_1$, as was the case in most of our runs, $\omega_-$ is given to good approximation ($\leq$ several percent error) by $\omega_{Heff}$ or $\omega_{Heff*}$ itself.

During one final run made with the more-massive lower diaphragm, data were also collected in the region of the intermediate angular frequency $\omega_+$. In addition, the value of $\omega_+$ was used in all of our runs in the determination of cell parameters. For a completely-filled cell, this angular frequency is given in the low-temperature limit to very good approximation by

$$\omega_+^2 \cong (1+\alpha_1)\omega_1^2 \left(1 + \frac{\alpha_1}{1+\alpha_1} \frac{\Omega_{2eff}}{\Omega_{1eff} + \Omega_{2eff}} \frac{\omega_{Heff}^2}{\omega_1^2}\right). \tag{12}$$

Here $\Omega_{2eff}$ (and $\omega_{Heff}^2$) are to be evaluated at $\omega_+$. Since we had $\omega_+ < \omega_2$ but not necessarily $\omega_+ \ll \omega_2$, the influence of $\omega_+^2/\omega_2^2$ on $\Omega_{2eff}$ and $\omega_{Heff}^2$ could not be neglected. However, when $\omega_{Heff} \ll \omega_1$, the dependence of the right-hand side of Eq. (12) on $\omega_+$ is weak and $\omega_+^2$ is given to good approximation by $(1+\alpha_1)\omega_1^2$ alone.

For a partially-filled upper chamber, we have

$$\omega_+^2 \cong (1+\alpha_1)\omega_1^2 \left(1 + \frac{\alpha_1}{1+\alpha_1} \frac{\omega_{Heff*}^2}{\omega_1^2}\right). \tag{13}$$

As in the completely-filled case, when $\omega_{Heff*} \ll \omega_1$, the quantity $\omega_+^2$ in the partially-filled case is given to good approximation by $(1+\alpha_1)\omega_1^2$ alone.

As described in the paragraphs below, we proceeded to use this model to determine phase-difference and velocity amplitudes from our measurements of response amplitude as follows. The amplitude of the response of the back-diode oscillator and fm detection circuit was used to determine the amplitude of the oscillation of the lower diaphragm. The model was then used to relate this amplitude to the amplitude of oscillation of the chemical potential difference between chambers. From the chemical potential difference amplitude we calculated the amplitude of oscillation of the order-parameter phase difference between chambers. This phase-difference amplitude could then be used to calculate the amplitude of the superfluid velocity oscillation in the small aperture. However, because of uncertainty in the constant of proportionality between the phase difference and the superfluid velocity, we report most of our results in terms of phase difference rather then in terms of superfluid velocity.

The amplitude of the oscillations of the chemical potential difference between chambers is given in general by

$$\mu_{20} - \mu_{10} = \frac{1}{\rho}(P_{20} - P_{10}) - s(T_{20} - T_{10}). \tag{14}$$

Here $\mu$ is the chemical potential difference per unit mass of the fluid, $s$ is the entropy per unit mass, and $T$ is the absolute temperature. In this formula and the following ones, the zero subscripts denote complex amplitudes associated with $exp(i\omega t)$ time dependences. For a completely-filled cell in the low-temperature limit, the chemical potential difference amplitude at resonance is given in terms of the amplitude of oscillation of the lower diaphragm to very good approximation by the expression

$$\mu_{20} - \mu_{10} \cong \frac{k_1}{\rho A_1} \frac{\Omega_{1eff} + \Omega_{2eff}}{\Omega_{2eff}} \left(1 - \frac{1}{1+\alpha_1}\left(1 + \frac{\alpha_1 \Omega_{2eff}}{\Omega_{1eff} + \Omega_{2eff}}\right)\frac{\omega^2}{\omega_1^2}\right)x_{10}. \tag{15}$$

This formula is applicable at subcritical levels of excitation, where we can assume linear behavior and a sinusoidal response to a sinusoidal drive, up to the critical level.

When the upper chamber is only partially filled, Eq. (15) is replaced by the simpler relationship

$$\mu_{20} - \mu_{10} \cong \frac{k_1}{\rho A_1}\left(1 - \frac{\omega^2}{\omega_1^2}\right)x_{10}. \tag{16}$$

When the entire temperature range that was studied experimentally is considered, two-fluid and thermal effects must be taken in account. However, as detailed in the Appendix, we find, somewhat to our surprise, that to an accuracy on the order of 1%, the only modification that must be made to the formulas given above is to replace the inductance $L$ that enters the expressions for $\omega_{Heff}^2$ and $\omega_{Heff*}^2$ with an effective inductance $L_{eff}$ that includes dissipative as well as inertial effects. In the limit that the normal fluid component is effectively immobilized by its viscosity in a small aperture, $L_{eff}$ is approximately equal to $(\rho/\rho_s)L$ and thus is strongly temperature-dependent. In the limit that the motion of the normal fluid component is influenced very little by viscosity in a large aperture, $L_{eff}$ is approximately equal to $L$ itself and has little temperature dependence. Hence for all of our data at $\omega_-$, chemical potential difference amplitudes were simply calculated using either Eq. (15) or (16) as appropriate, setting $\omega$ equal to the experimental resonant angular frequency. For the filled-cell data collected at $\omega_+$ in the final run, Eqs. (12) and (15) were combined to yield

$$\mu_{20} - \mu_{10} \cong -\frac{k_1}{\rho A_1} \alpha_1 x_{10}, \qquad (17)$$

to good approximation.

The amplitude of the order-parameter phase-difference oscillations was calculated from the chemical potential difference amplitude using the Josephson-Anderson equation[10,21] in the form

$$\phi_{20} - \phi_{10} = -\frac{m_4}{i\omega\hbar}(\mu_{20} - \mu_{10}). \qquad (18)$$

Here $\phi$ is the phase of the superfluid order parameter, $m_4$ is the mass of the $^4$He atom, and $\hbar$ is Planck's constant divided by $2\pi$.

Finally, we can estimate the amplitude of the average superfluid velocity in the aperture, $v_{s0}$, from the phase-difference amplitude using the relation

$$v_{s0} = \frac{\hbar}{m_4 \ell_{eff}}(\phi_{20} - \phi_{10}). \qquad (19)$$

Here $\ell_{eff}$ is the effective fluid-dynamical length of the aperture, as defined by Eq. (5), applied to the superfluid component.

The parameters appearing in Eqs. (15), (16), and (17) were determined as follows. The quotient $k_1/A_1$, in combination with the response of the back-diode oscillator frequency to the displacement of the lower diaphragm, was determined statically at 0.60 K by applying known pressure changes to the interior of the cell from outside the cryostat through the fill capillary and measuring the resulting changes in the back-diode oscillator frequency. This combination was assumed to be temperature-independent. The quantities $\alpha_1$ and $(\Omega_{1eff} + \Omega_{2eff})/\Omega_{2eff}$ were estimated at 0.60 K from the observed values of $\omega_1$ with an empty cell and of $\omega_-$ and $\omega_+$ with the cell partially and completely filled, using Eqs. (6), (7), and (10)-(13). In determining these parameters and applying them at other temperatures, variations of $\rho$ and $\kappa$ with temperature and pressure were taken into account.[22,23]

Various discrepancies appeared between the model and the experiment. Rather than being the same from run to run with the same lower diaphragm, the values of $\alpha_1$ and $(\Omega_{1eff} + \Omega_{2eff})/\Omega_{2eff}$ (at $\omega_-$), as determined above, varied more than might have been expected merely from the disassembly and reassembly of the cell to change the bypass channel. As additional checks of the consistency of the model, estimates of the fluid-dynamical inductances of the small aperture and of the bypass channels were made at 0.60 K, based on geometrical considerations. It was found that a consistent set of inductance values could be constructed in relation to the observed resonant frequencies if allowance was made for additional distributed inductance within the lower chamber, inductance which resulted from the narrowness of that chamber and which became important when the largest bypass channels were used to obtain the highest values of $\omega_-$.[17] These inductances, together with the measured partially-filled resonant angular frequencies $\omega_-$, yielded values of $\Omega_{1eff}$ averaging $(1.24 \pm 0.13) \times 10^{-6}$ m$^3$ for several runs with the less-massive diaphragm. For the same runs, the average value of $\alpha_1$ was $2.16 \pm 0.27$, yielding a value for $\Omega_1$ of $(0.39 \pm 0.06) \times 10^{-6}$ m$^3$, slightly more than twice the estimated geometrical value of $0.17 \times 10^{-6}$ m$^3$. On the other hand, an independent calculation of $\alpha_1$ for the less-massive diaphragm, based on the dimensions and material characteristics of the diaphragm, produced a value of 5.2, which yields a value for $\Omega_1$ of $0.20 \times 10^{-6}$ m$^3$, in reasonable agreement with the estimated geometrical value.

It should be noted that for the measurements of phase-difference amplitude at $\omega_-$ based on Eq. (15), the results are rather insensitive to $\alpha_1$ at the lower values of $\omega_-$. In addition, although these results are nearly proportional to $(\Omega_{1eff} + \Omega_{2eff})/\Omega_{2eff}$, this factor is almost independent of temperature and frequency at $\omega_-$. Therefore, unless the model has some serious unsuspected inadequacy, the inconsistencies noted above should not influence the temperature dependence of the results at $\omega_-$ significantly nor the comparison of results from run to run. However, the absolute magnitudes of the results might be in some error. Neither of these quantities influences results based on Eq. (16) for the partially-filled case.

In the runs with the less-massive diaphragm, there is some reason to doubt our determinations of $\omega_+$ and thus our determinations of $\alpha_1$ from $\omega_+$. In these runs, the apparent value of $\omega_+$ of $\sim 2\pi \times 8000$ rad/s lay between the estimated angular frequencies of the first and second first-sound modes in the chambers. Some coupling to these modes, not included in the model, may have shifted the resonance frequency. In the run with the more-massive diaphragm, the resonance at $\omega_+$ near $2\pi \times 1900$ rad/s, although much lower than the lowest first-sound resonance angular frequency, was split into components at $2\pi \times 1884$ and $2\pi \times 1914$ rad/s at 0.60 K for some unknown reason, thus casting doubt on the determination of $\alpha_1$ in this run too.

For the phase-difference results at $\omega_+$ based on Eq. (17) in the run with the more-massive diaphragm, there is much more sensitivity to $\alpha_1$ than for the results at $\omega_-$ based on Eq. (15), although the dependence on $(\Omega_{1eff} + \Omega_{2eff})/\Omega_{2eff}$ is absent. Thus there is also some question about the absolute value of the results at $\omega_+$, although the temperature dependence of these results should be reliable. Our analysis was based on the more-prominent lower-frequency component of the resonance.

## IV. RESULTS

Critical-superfluid-flow behavior was studied as a function of temperature in a total of seven experimental runs, as listed in the Table. The first six of these were made with the less-massive lower diaphragm. These six involved two runs with the small aperture alone, followed by four runs at successively higher frequencies as a series of increasingly-open bypass channels was installed. The seventh run was made with the more-massive lower diaphragm, with the small aperture alone connecting the chambers of the cell. This last run permitted us to check whether the same change in critical-velocity behavior that was observed to occur at the lower temperatures with increasing frequency in the first six runs could be seen with the small aperture alone in a single run, by making observations at both $\omega_-$ and $\omega_+$. For measurements with a completely-filled cell, the pressure was usually greater than saturated vapor pressure but was always less than 0.5 bar.

TABLE. Parameters for the various sets of critical phase-difference data.

| Run number | Filling | Frequency at 0.60 K (Hz) | Frequency at 2.00 K (Hz) | Normalization factor | Avalanche onset T (K) |
|---|---|---|---|---|---|
| 1 | Complete | 73 | 47 | 1.00 | --- |
| 2 | Complete | 73 | 47 | 1.00 | --- |
| 3 | Partial | 120 | | 0.82 | --- |
| 3 | Complete | 186 | | 0.97 | --- |
| 4 | Partial | 320 | 313 | 0.87 | --- |
| 4 | Complete | 505 | 483 | 0.98 | 0.6 < T < 0.7 |
| 5 | Partial | 653 | 643 | 0.87 | 0.8 < T < 0.9 |
| 5 | Complete | 969 | 935 | 1.00 | 1.1 < T < 1.2 |
| 6 | Partial | 1036 | 1025 | 0.95 | 0.9 < T < 1.0 |
| 6 | Complete | 1448 | 1397 | 1.06 | 1.0 < T < 1.2 |
| 7 | Partial | 83 | 53 | 0.76 | --- |
| 7 | Complete | 1884 | 1848 | 0.63 | 1.0 < T < 1.1 |

In Fig. 3 we show the critical phase difference as a function of temperature determined in the lower range of frequencies studied. Included are data determined at $\omega_-$ with a completely-filled cell in the first three runs and with a partially-filled cell in the third and fourth runs. The values of $\Delta\phi_c/2\pi$ have been arbitrarily normalized by the factors listed in the Table to bring the various curves into good coincidence, using data from the first two runs as a reference. As mentioned earlier, the critical velocity $v_c$ is proportional to $\Delta\phi_c/2\pi$. Using Eq. (19) with $\ell_{eff} = 1.7$ μm, we estimate that a value of $\Delta\phi_c(\text{in radians})/2\pi = 25$ corresponds approximately to a value of $v_c = 1.5$ m/s.

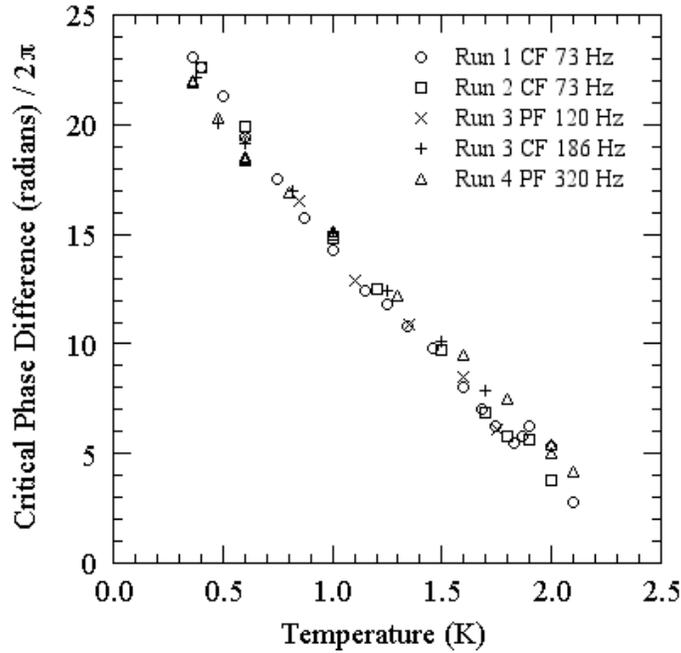

FIG. 3. Normalized critical phase difference versus temperature in the lower range of frequencies studied, from the first four runs. The symbols are identified by run number and "PF" or "CF" specifying whether the cell was partially or completely filled. Frequencies are listed for 0.60 K.

In Fig. 4 we show critical-phase-difference data at $\omega_-$ from the fourth, fifth, and sixth runs with a completely-filled cell, and from the fifth and sixth runs with a partially-filled cell. These data have also been normalized by factors included in the Table. In this case, the factors were chosen both to bring these data into the best possible coincidence with each other and to agree with the data in Fig. 3 at 0.60 K. The data plotted in both Figs. 3 and 4 were determined using values of $\alpha_1$ and $(\Omega_{1eff} + \Omega_{2eff})/\Omega_{2eff}$ (at $\omega_-$) averaged over several of the first six runs.

In Fig. 5 we show critical-phase-difference data determined in the last run at $\omega_-$ and at $\omega_+$. Here the normalization factor for the data at $\omega_-$ was chosen to bring these data into agreement with the data of Figs. 3 and 4 at 0.60 K, and the factor for the data at $\omega_+$ was chosen to bring those data into agreement with the data at $\omega_-$ in this figure at 1.3 K and above.

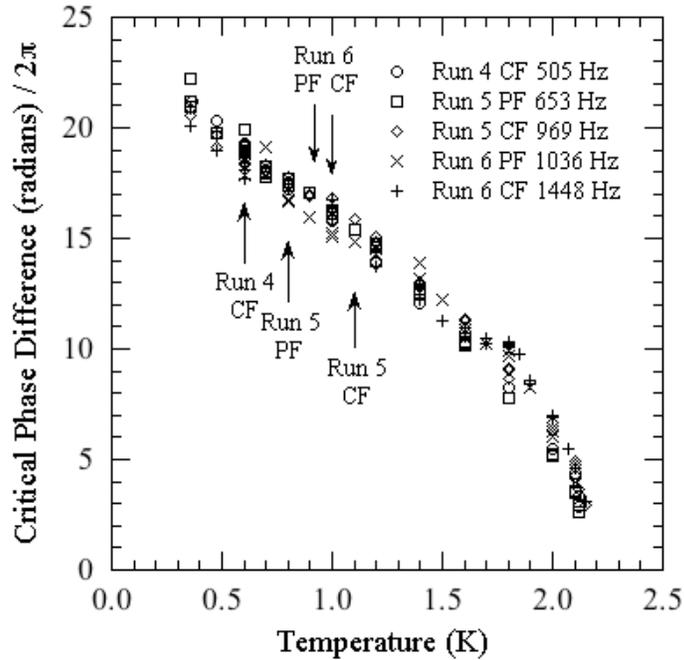

FIG. 4. Normalized critical phase difference versus temperature in the higher range of frequencies studied, from the fourth, fifth, and sixth runs. The symbols are identified by run number and "PF" or "CF" specifying whether the cell was partially or completely filled. Frequencies are listed for 0.60 K. The vertical arrows mark the highest temperatures at which large energy-loss events were resolved in each data set.

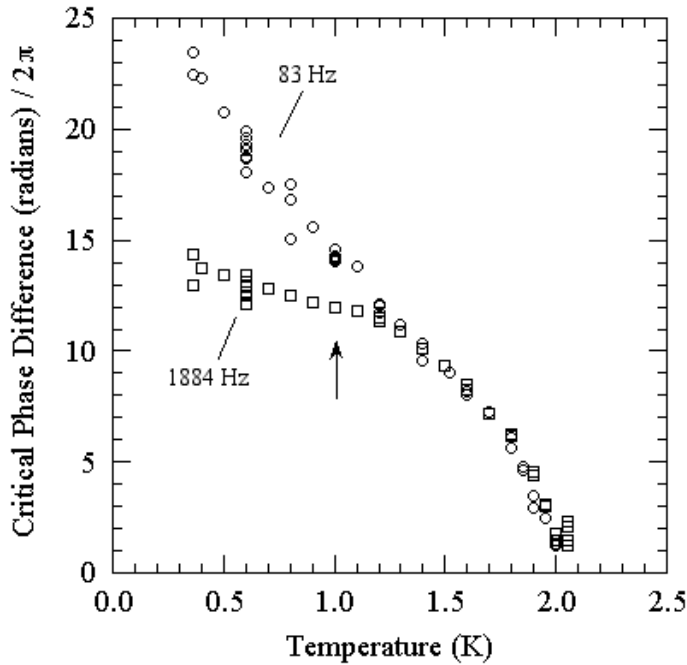

FIG. 5. Normalized critical phase difference versus temperature from the last run, showing data collected at $\omega_-$ (circles) and at $\omega_+$ (squares). Frequencies are listed for 0.60 K. The cell was completely filled for both sets of data. The vertical arrow marks the highest temperature at which large energy-loss events were resolved at $\omega_+$.

For all of the data plotted in Fig. 3 and for the data at $\omega_-$ in Fig. 5, the supercritical response amplitude was a smooth and monotonically increasing function of drive amplitude, with relatively minor fluctuations, as illustrated in Fig. 6. In the runs with the small aperture alone, in particular the second and the last runs, the response as a function of time at drive levels that were just barely supercritical was seen to take the form of a relatively regular sawtooth pattern, characteristic of $2\pi$ phase slips separated by intervals of recovery. An example of this response is shown in Fig. 7. Phase-slip events larger than $2\pi$ phase slips were extremely rare.

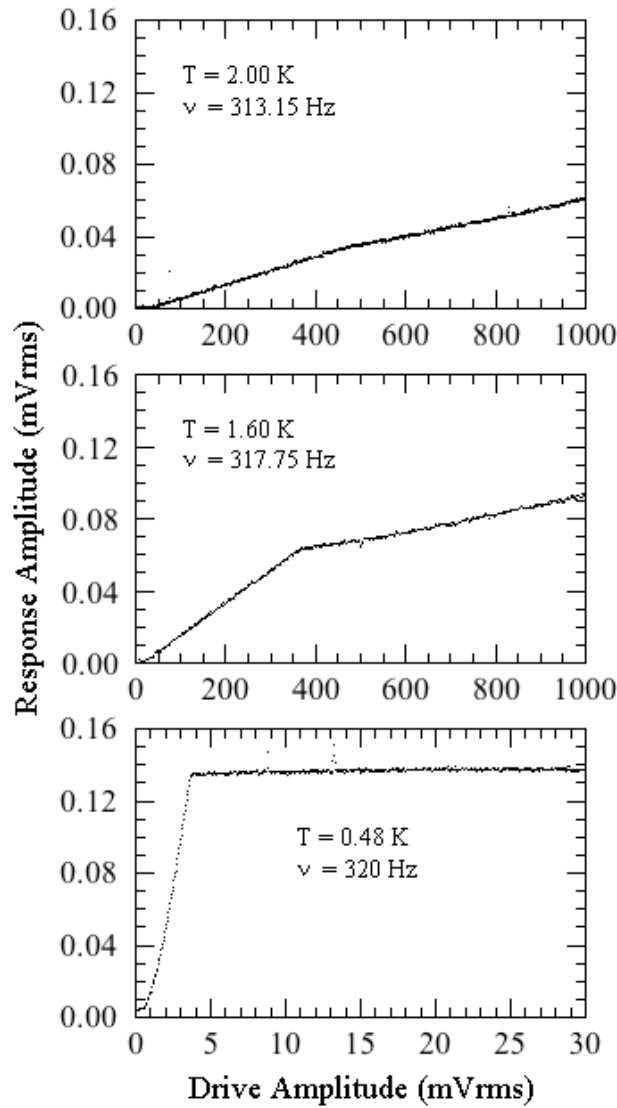

FIG. 6. Response amplitude versus drive amplitude at three representative temperatures from the fourth experimental run with a partially-filled cell. This case provided the highest-frequency data that did not exhibit large collapses.

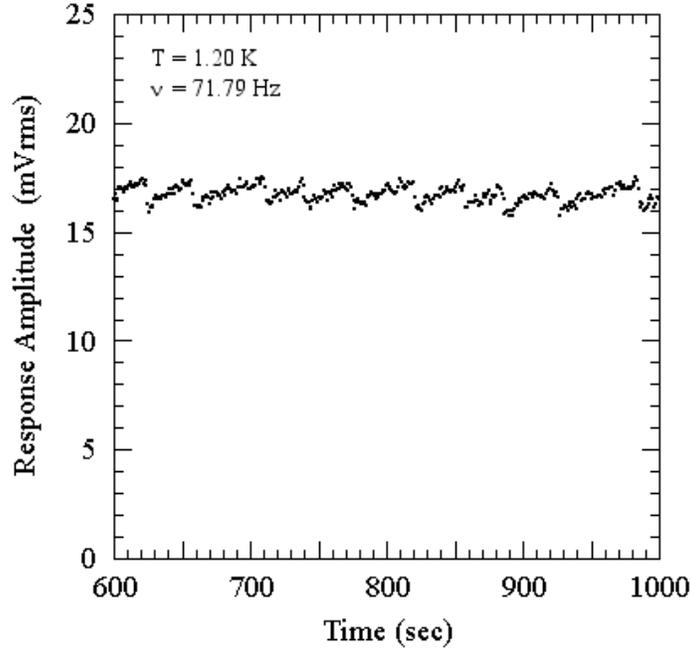

FIG. 7. Response amplitude versus time measured at a slightly supercritical drive level during the second run, showing a series of abrupt energy-loss events believed to be single $2\pi$ phase slips.

If these phase slips are indeed "full" $2\pi$ phase slips, reflecting the complete crossing of the potential-flow streamlines by individual independent vortices at the maximum flow rate, the loss of amplitude due to one slip provides an independent check on the calibration of the response of the cell.[9,21,24,25] From phase slips observed in the second run from 0.4 to 1.5 K, we would infer that the values of critical phase difference determined by the method of calibration described in Section III are approximately 30% too small.

In the data from the third and fourth runs represented in Fig. 3, in which bypass channels were used, no such sawtooth fluctuations were observed. Although $2\pi$ phase slips at the small aperture were presumably still responsible for the supercritical behavior, the presence of the bypass channel would have reduced their influence on the amplitude of the phase difference between reservoirs to the point where they could no longer be resolved.

For the runs represented in Fig. 4, the same apparent supercritical behavior as above was seen at the higher temperatures. In all cases, the $2\pi$ phase slips that were presumably responsible for supercritical behavior would have been too small to be resolved. However, for all of these sets of data, a marked change in supercritical behavior was noted at the lower temperatures. At drive levels just above the onset of critical behavior, erratic events occurred spontaneously that involved relatively rapid losses of energy stored in the resonator followed by periods of relatively slow recovery at the rate expected for noncritical behavior. As the drive level increased, the rate of recovery increased, as did the frequency of occurrence of these events, until they were no longer resolvable from each other.

The average loss of energy in each of these irregular events increased gradually as the temperature decreased below some onset temperature. This behavior is illustrated in Fig. 8, and was seen with the cell both completely and partially filled. The vertical arrows in Figs. 4 and 5 show these onset temperatures for the various sets of data, temperatures that are also listed in the Table. The critical amplitudes plotted at temperatures below these onset temperatures represent average response amplitudes at drive levels just barely supercritical, in some cases in the presence of rather large fluctuations.

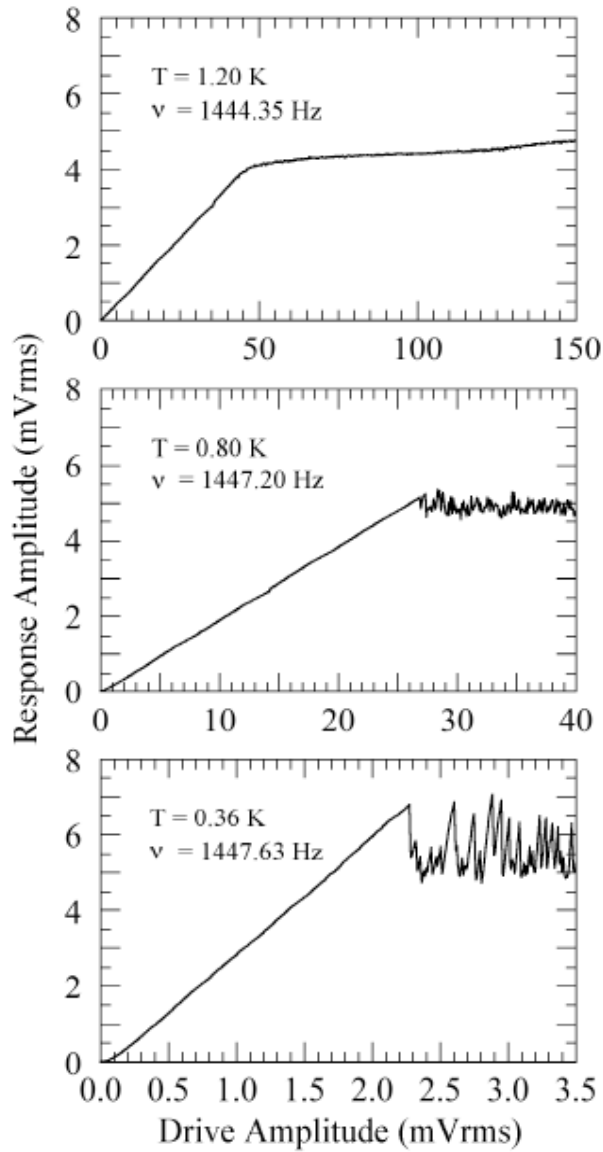

FIG. 8. Response amplitude versus drive amplitude at three representative temperatures from the sixth run with a completely-filled cell. The two lower curves show the large energy-loss events that appeared in the supercritical region at lower temperatures for frequencies of 500 Hz and above.

The large energy-loss events seen at the higher frequencies, at the lower temperatures, differ from $2\pi$ phase slips in a number of ways. A detailed picture of several such events is shown in Fig. 9, recorded during the sixth run with a completely-filled cell at a temperature of 0.36 K and a resonant frequency of 1448 Hz. See also Refs. 16 and 17. These events represented relatively abrupt energy losses equivalent to the energy that would be dissipated by 10's or 100's of individual $2\pi$ phase slips at maximum potential flow. Furthermore, the distribution of values of phase difference from which they occurred was much broader than that for $2\pi$ phase slips. The distribution of the amounts of energy lost was also very broad, unlike that for the $2\pi$ phase slips, which appeared to be a narrow peak within our experimental resolution. Although relatively abrupt, the energy loss in these events was observed to occur over periods of many cycles of oscillation (see e.g. the caption to Fig. 9), indicating some sort of sustained dissipation process. By contrast, although beyond the resolution of our current experiment, individual $2\pi$ phase slips are thought to require at most a few cycles for completion at this frequency.[26] Thus it seems appropriate to refer to the large energy-loss events as avalanches.

The same transition in behavior from single $2\pi$ phase slips to avalanches was also observed in the higher-frequency data obtained from the final experimental run. We see from Fig. 5 that with suitable normalization, the critical-phase-difference data from both $\omega_-$ and $\omega_+$ above 1.3 K can be brought into good agreement. However, below 1.3 K, in the region where large energy-loss events are present at $\omega_+$, the two sets of critical-phase-difference data have different temperature dependences, with the data at $\omega_+$ exhibiting less overall temperature dependence and lying below the data at $\omega_-$ as presently normalized. This behavior is reminiscent of the behavior seen in an earlier experiment using this same apparatus but with a different aperture and with a bypass channel present.[15] The data from our final run help to reassure us that the appearance of large energy-loss events with increasing frequency is not an artifact associated with the presence of a bypass channel, nor does it reflect some progressive change in the properties of the aperture during the course of the present work. When all of our data are considered together, the onset temperature for avalanches appears to increase with increasing frequency.

Our most extensive set of avalanche data was collected during the sixth run with a completely-filled cell at a temperature of 0.601 K and a resonant frequency of 1448 Hz. These data are representative of avalanche data collected under other conditions. In Fig. 10 we plot the final response amplitude versus the initial amplitude for each of 753 avalanches. Both the initial and final values are widely distributed over ranges that are approximately 15% wide relative to the average amplitude, which was taken to be the effective critical amplitude for these data. The initial value has rather well-defined lower and upper bounds, and the final value has a rather well-defined lower bound. The initial and final values are not noticeably correlated, save for the condition that the final value cannot be greater than the initial value. The distribution of initial values is a broad peak favoring smaller values. The distribution of final values is a slightly narrower peak, also favoring smaller values.

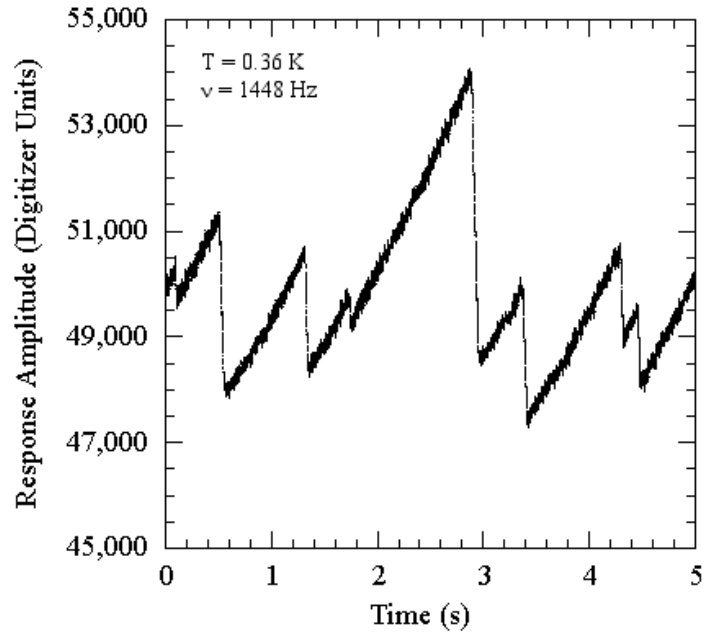

FIG. 9. Representative plot of the response amplitude versus time measured at a low supercritical drive level during the sixth run, showing the details of several large energy-loss events (avalanches). These data were recorded with a completely-filled cell at $v_- = 1448$ Hz and $T = 0.36$ K. One thousand digitizer units correspond to a phase difference in radians divided by $2\pi$ of approximately 0.38. The duration of the largest avalanche, occurring at 2.9 s, was 67 ms or approximately 97 cycles of oscillation. The energy dissipated in this avalanche was approximately 700 times the energy that would have been dissipated by a single $2\pi$ phase slip in dc flow at the maximum rate of ac flow through the small aperture. The average rate of dissipation corresponds to the loss of one-third of the maximum kinetic energy of flow through the small aperture every half-cycle of oscillation.

The data in Fig. 10 were collected, in fact, at four different supercritical drive levels, ~ 2.5, 5, 10, and 20 times the critical drive level. However, most of the data come from measurements at ~ 5 and 20 times the critical drive level. The separate distributions for these two widely differing drive levels are almost identical. However, for drive levels appreciably greater than the higher of these two levels, the distributions are altered, the end values tending to rise, as an appreciable amount of energy is supplied by the drive during the course of an energy-loss event and the duration of such an event is lengthened.

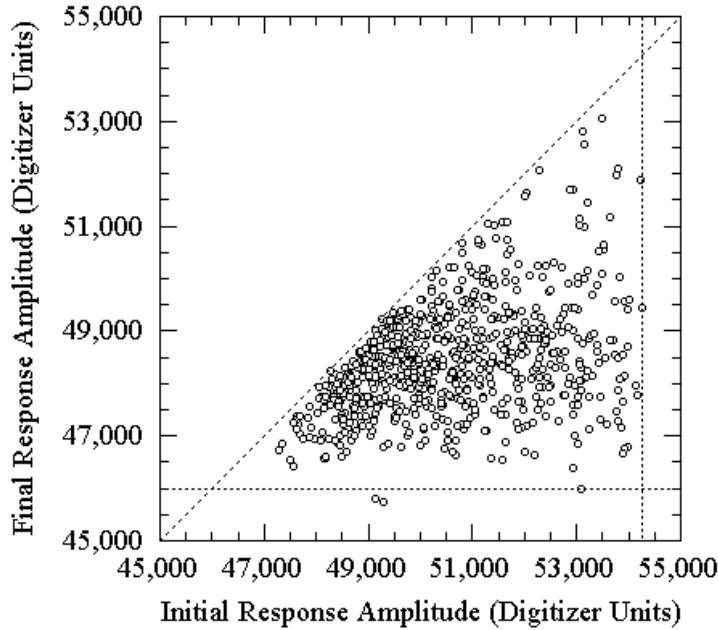

FIG. 10. Final response amplitude versus initial response amplitude for a total of 753 large energy-loss events (avalanches) measured with a completely-filled cell at 0.60 K during the sixth run. The frequency was 1448 Hz. The same conversion between digitizer units and phase difference divided by $2\pi$ as in Fig. 9 applies here. The diagonal dashed line, along which initial and final amplitudes are equal, provides an absolute bound for the data: all events must lie below and to the right of this line. The vertical and horizontal dashed lines are merely guides to the eye emphasizing the effective bounds of the observed initial and final response amplitudes, respectively.

There has been much recent attention given to avalanche-like behavior in a variety of systems, its possible universality, and its possible relation to self-organized criticality. This behavior includes sand and rice-pile avalanches, earthquakes, droplet formation, pulsar glitches, solar flares, and vortex avalanches in superconductors.[27,28] Thus it is of interest to characterize the avalanches observed here in a way that allows comparison with those in other systems.

One possible comparison concerns the distribution of avalanche size, in our case the drop in response amplitude that occurs during an avalanche. Fig. 11 shows a log-log plot of a histogram of the loss of response amplitude occurring during the events shown in Fig. 10 using a bin-width of 200 digitizer units. We see that over the decade and a half of avalanche size observed, there is no sign of the inverse power-law behavior that is seen in a variety of other physical systems exhibiting avalanche-like behavior and that is expected for scale-independent self-organized criticality. In comparison to such behavior, our results are deficient in both the largest and the smallest events. Our results decrease monotonically with increasing size and can be fitted rather well with a simple ad hoc quadratic form, as shown in the figure.

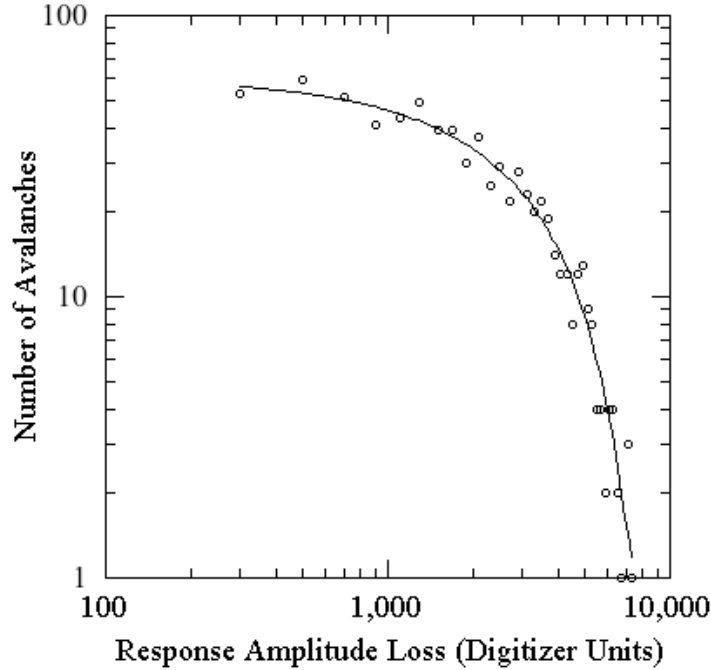

FIG. 11. Log-log plot of a histogram of the loss of response amplitude occurring during the events shown in Fig. 10. The circles were calculated using a bin-width of 200 digitizer units. The smooth curve is merely an ad hoc quadratic fit to the data having the form $N = 60.51 - 154.2 \times 10^{-4} L + 99.84 \times 10^{-8} L^2$, where $N$ is the number of avalanches and $L$ is the response amplitude loss in digitizer units.

Another possible comparison concerns the power spectrum of the avalanches. In the analysis of other avalanche experiments, it seems usual to consider time-series that are analogs of the time-derivative of our response amplitude, which would be an irregular series of negative spikes on a nearly-constant positive background. See, for example, Refs. 29-31. Thus, for purposes of comparison, we have calculated power spectra of the derivative of our signal, by multiplying the Fourier transform of the signal by the frequency and squaring the absolute value of the product. Fig. 12 is a log-log plot of such a power spectrum for a 20-s-long time series of response amplitudes.

In order to distinguish the avalanche-related part of the power spectrum from instrumental noise and interference, the open circles in Fig. 12 were calculated from a "clean" version of the time series, constructed by connecting successive beginning and end points of the 64 avalanches present in the data with straight lines. The dots in the figure were calculated from the raw data. A comparison with the power spectrum of data recorded at a subcritical drive amplitude, in which avalanches are absent and only noise is present, shows that the divergence of the dots from the circles in Fig. 12 above approximately 40 Hz represents the avalanche contribution to the spectrum being overwhelmed by noise. The plots have been terminated somewhat arbitrarily at 200 Hz.

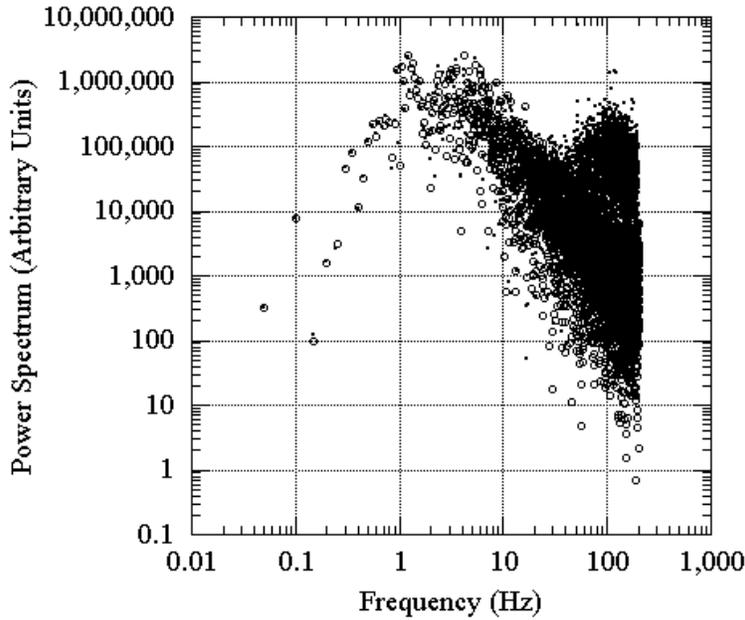

FIG. 12. Log-log plot of the power spectrum of the time-derivative of the response amplitude for 20 s of avalanche data, recorded under the same conditions as the data in Fig. 9 except for a drive level three times larger. There were 64 avalanches in this time interval. The dots show the results from the raw data. The open circles show the results from "clean" data, constructed by connecting successive initial and final points of avalanches with straight lines to minimize the influence of noise. Notice that the dots and circles begin to diverge above approximately 40 Hz, as noise and interference in the raw data begin to dominate the signal.

We see that the avalanche part of the spectrum consists of a little over one decade of rising behavior at the lowest frequencies, approximately proportional to $f^3$, where $f$ is the frequency, followed by a little over one decade of falling behavior, approximately proportional to $f^{-2}$. This sequence of behavior is typical of the behavior at other drive levels and temperatures with the following qualifications. The rise that is approximately proportional to $f^3$ at the lowest frequencies is more generally proportional to $f^m$ with $m$ in the range from 2 to 3. This rise is almost absent at low drive levels and becomes more pronounced and extended in frequency with increasing drive level, as the number of avalanches per unit time increases. The rise appears to end at a frequency approximately equal to the frequency with which avalanches would occur if all of them were equal in size to the largest avalanches observed. Thus the peaked spectrum reflects the existence of a characteristic frequency for the avalanches. The fall that occurs at frequencies above the maximum in the spectrum, and that is approximately proportional to $f^{-2}$ in the figure, is more generally proportional to $f^n$ with $n$ in the range from -1 to -2.

Thus any evidence in our data of the $f^{-1}$ behavior seen in some other systems and expected for self-organized criticality, is rather meager. In so far as the maximum size of our avalanches is limited and that, as a result, there is a moderately well-defined maximum recovery time, our avalanches are similar to the sandpile avalanches of Jaeger, Liu, and Nagel, who failed to see a $f^{-1}$ region.[29]

## V. CONCLUSIONS AND DISCUSSION

Our measurements of critical-velocity behavior in the oscillatory flow of superfluid $^4$He through a 2-µm by 2-µm aperture in a 0.1-µm-thick foil have shown that a transition of critical-velocity behavior occurs with increasing frequency of oscillation. For observations at frequencies below 400 Hz made during the early runs, the critical phase difference and critical velocity were nearly linearly-decreasing functions of increasing temperature over the entire temperature range studied, from 0.36 to 2.10 K, extrapolating to zero at ~ 2.4 K and to a critical phase difference of ~ 26 × 2π radians at 0 K. This temperature dependence is typical of that observed for apertures with at least one submicron linear dimension transverse to the flow, but such a temperature dependence is less well documented for apertures as open as the present one.[32-34] However, some curvature is often seen, the temperature to which the critical phase difference extrapolates to zero can vary by a few tenths of a kelvin, and the critical phase difference extrapolated to 0 K can vary by as much as an order of magnitude.[8,9,11,15,19,35]

For observations made at frequencies above 400 Hz in the later runs, downward curvature was observed in the decrease of critical velocity with increasing temperature, at the higher temperatures. Furthermore, a pronounced alteration in supercritical-flow behavior was observed at the lower temperatures. The temperatures at which this altered behavior was observed rose from ~ 0.6 K and below, at 500 Hz, to ~ 1.0 K and below, at 1880 Hz. Downward curvature was also seen at low frequency in the last run, at the higher temperatures. This result might suggest that the downward curvature observed in the later part of our work resulted from some change in the aperture during the series of experimental runs, rather than from the increases in frequency. Nevertheless, the results of the last run confirmed the dependence on frequency of the altered supercritical-flow behavior seen in the later runs at the lower temperatures.

At the lower frequencies at all temperatures, and at the higher frequencies in the higher-temperature region, the onset of dissipation was consistent with the appearance of single 2π phase slips involving the nucleation and growth of individual vortex loops. Such phase slips could be observed directly at the very lowest frequencies studied. At the higher frequencies in the higher-temperature region, even though individual phase slips could not be resolved, the smooth and reproducible supercritical response observed was consistent with this mechanism of dissipation. A critical velocity that decreases with increasing temperature in a nearly linear fashion can be understood in terms of a model that combines thermal activation with fluid-dynamic growth of superfluid vortices.[2,4,5,8,9,11,36]

On the other hand, in the region of altered behavior at the higher frequencies and lower temperatures, the onset of dissipation was dominated by large, irregular energy-loss events or avalanches. Typical avalanche events involved a rate of energy loss equivalent to the production of a number of isolated vortex loops per half-cycle of oscillation, and lasted from fewer than 10 to more than 100 cycles of oscillation. The average rate of energy loss during avalanches at 0.36 K during Run 6, as illustrated in Fig. 9, corresponded to the loss of approximately one-third of the maximum kinetic energy of flow through the small aperture every half-cycle of oscillation, whereas during avalanches at 0.60 K in Run 5, it corresponded approximately to the complete loss of the maximum kinetic energy of flow through the small aperture every half-cycle.[16,17] The present work supports and extends earlier observations of a similar alteration of supercritical-flow behavior in an aperture in the shape of a narrow slot 5 μm long and 0.3 μm wide in a foil 0.2 μm thick.[15] It is noteworthy, nevertheless, that avalanche dissipation was not observed in previous work with a 0.2-μm-diameter aperture in a 0.1-μm-thick foil at frequencies up to 2850 Hz.[19]

As is discussed in a separate article,[26] a simple model of vortex motion near an aperture in the presence of oscillatory flow suggests that this altered behavior may result from the drawback of vortices into the aperture, where they are reused for further energy loss without the need for further nucleation, in some sort of "vortex mill". Such a process would be favored by higher frequencies of oscillation. However, the details of how such a process might lead to the observed high rate of energy loss that can persist for a number of half-cycles remain unclear.

It is tempting to consider whether the transition in behavior observed here and in the earlier work mentioned above reflects a transition from the large and highly temperature-dependent critical velocities typically seen in submicron-size apertures, to the small and weakly temperature-dependent critical velocities often seen in apertures and longer passages having diameters of 10 microns and larger.[2] This suggestion would envisage the frequency of the transition dropping to zero with increasing aperture size. However, the critical velocities in the higher-frequency region in this work and the earlier work are still relatively large and show considerable temperature dependence, although this dependence is less than in the lower-frequency region. Thus any connection of the higher-frequency behavior with Feynman-like critical velocities in larger channels is still uncertain and remains an interesting topic for investigation.

## ACKNOWLEDGEMENTS

We wish to acknowledge the support of this work by the National Science Foundation through grants DMR 90-02890, 94-03522, and 96-31703. In addition, we would like to thank James Kakalios for discussions regarding avalanche behavior.

## APPENDIX

In this appendix we consider a model for the operation of our cell, from which we derive the relationships used in interpreting our data. The present model is a generalization of the one used by Beecken and Zimmermann to treat the same cell, this time including the compliance and inertia of the upper diaphragm.[18,19] Similar modeling has been done by Backhaus and Backhaus for a somewhat different cell with somewhat different emphasis.[37]

The model for our cell is shown schematically in Fig. 13. The lower and upper chambers are designated 1 and 2, respectively. The symbols $I_s$ and $I_n$ represent, respectively, the superfluid and normal fluid mass currents between chambers, positive in the upward direction, and $x_1$ and $x_2$ represent, respectively, the displacements of the centers of the lower and upper diaphragms, positive in the upward direction. In this model we restrict our attention to subcritical superfluid flow and, when two channels between chambers are present, we assume that no changes in superfluid circulation involving the two channels take place, so that the two apertures in parallel can be treated as a single flow channel. First we consider a cell completely filled with liquid.

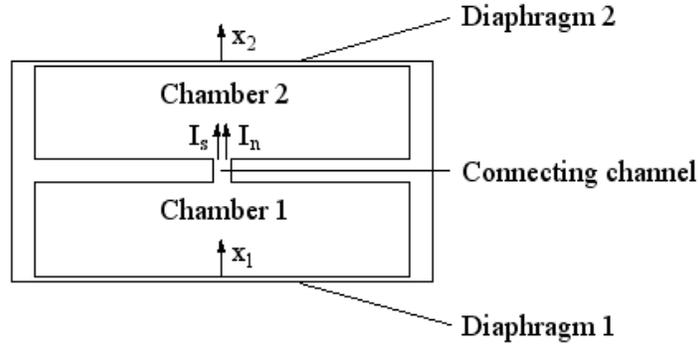

FIG. 13. Schematic drawing of the cell.

The first two equations below express the conservation of mass in the first and second chambers, respectively.

$$I_s + I_n = -\left(\frac{\partial \rho}{\partial P}\right)_T \Omega_1 \frac{dP_1}{dt} - \left(\frac{\partial \rho}{\partial T}\right)_P \Omega_1 \frac{dT_1}{dt} + \rho A_1 \frac{dx_1}{dt} , \qquad (A1)$$

$$I_s + I_n = \left(\frac{\partial \rho}{\partial P}\right)_T \Omega_2 \frac{dP_2}{dt} + \left(\frac{\partial \rho}{\partial T}\right)_P \Omega_2 \frac{dT_2}{dt} + \rho A_2 \frac{dx_2}{dt} . \tag{A2}$$

Here $\rho$ is the mass per unit volume of the fluid, $\Omega_1$ and $\Omega_2$ are the volumes of the chambers, $P_1$ and $P_2$ are the pressures in the chambers, $T_1$ and $T_2$ are the temperatures in the chambers, $A_1$ and $A_2$ are the effective areas of the diaphragms, and $t$ is the time. Each effective area is defined such that its product with the corresponding $x$ equals the volume displaced by the diaphragm.

The next two equations express the (approximate) conservation of entropy in the first and second chambers, respectively.

$$\frac{\rho s}{\rho_n} I_n - \frac{T_2 - T_1}{TR_T} = -\left(\frac{\partial(\rho s)}{\partial P}\right)_T \Omega_1 \frac{dP_1}{dt} - \left(\frac{\partial(\rho s)}{\partial T}\right)_P \Omega_1 \frac{dT_1}{dt} + \rho s A_1 \frac{dx_1}{dt} , \tag{A3}$$

$$\frac{\rho s}{\rho_n} I_n - \frac{T_2 - T_1}{TR_T} = \left(\frac{\partial(\rho s)}{\partial P}\right)_T \Omega_2 \frac{dP_2}{dt} + \left(\frac{\partial(\rho s)}{\partial T}\right)_P \Omega_2 \frac{dT_2}{dt} + \rho s A_2 \frac{dx_2}{dt} . \tag{A4}$$

Here $s$ is the entropy of the fluid per unit mass and $\rho_n$ is the normal fluid density. The quantity $R_T$ represents the total thermal resistance between chambers, including the Kapitsa resistance at the walls of each chamber and any thermal resistance in the walls between chambers, evaluated at an average temperature $T$. We assume for this purpose that the cell walls have negligible heat capacity and negligible thermal contact with the surroundings.

Next we write equations of motion for the lower and upper diaphragms, respectively.

$$m_1 \frac{d^2 x_1}{dt^2} = -k_1 x_1 - A_1 P_1 , \tag{A5}$$

$$m_2 \frac{d^2 x_2}{dt^2} = -k_2 x_2 + A_2 P_2 + \beta V . \tag{A6}$$

Here $m_1$ and $m_2$ are the effective masses of the diaphragms, defined such that $m(dx/dt)^2/2$ equals the kinetic energy of the diaphragm, $k_1$ and $k_2$ are the effective spring constants of the diaphragms, defined such that $kx^2/2$ equals the elastic potential energy of the diaphragm, and $A_1$ and $A_2$ can be shown to be the same effective areas as were defined in Eqs. (A1) and (A2). $V$ is the voltage applied to the piezoelectric element that drives the upper diaphragm, and $\beta$ is a constant coefficient relating $V$ to the effective drive force applied to the upper diaphragm.

The final two introductory equations, given below, represent equations of motion for the superfluid and normal fluid components, respectively.

$$\frac{\rho}{\rho_s} L_s \frac{dI_s}{dt} = -\frac{1}{\rho}(P_2 - P_1) + s(T_2 - T_1), \tag{A7}$$

$$\frac{\rho}{\rho_n} L_n \frac{dI_n}{dt} + \frac{\rho^2}{\rho_n^2} R_n I_n = -\frac{1}{\rho}(P_2 - P_1) - \frac{\rho_s s}{\rho_n}(T_2 - T_1). \tag{A8}$$

Here $\rho_s$ is the superfluid density. Equation (A7) follows from the usual superfluid equation of motion, together with the condition that the flow be potential flow. The quantity $(\rho/\rho_s)L_s$ is a fluid-dynamical inductance, defined such that $(\rho/\rho_s)L_s I_s^2/2$ equals the kinetic energy of the superfluid flow. For a single long cylindrical channel, $L_s = \ell/(\rho S)$, where $\ell$ is the length of the channel and $S$ is the cross-sectional area. Equation (A8) does not follow directly from the usual normal fluid equation of motion but is instead an interpolation formula between the inertial flow regime, in which the first term on the left hand side dominates, and the viscous regime, in which the second term dominates. The quantity $(\rho/\rho_n)L_n$ is completely analogous to $(\rho/\rho_s)L_s$, with $L_n = \ell/(\rho S)$ for a single long cylindrical channel. The quantity $(\rho^2/\rho_n^2)R_n$ is a fluid-dynamical resistance, defined such that $(\rho^2/\rho_n^2)R_n I_n^2$ is the rate of viscous energy loss in steady flow. For a single long circular cylindrical channel, $R_n = 8\eta_n \ell/(\pi \rho^2 a^4)$, where $a$ is the channel radius and $\eta_n$ is the coefficient of shear viscosity of the normal fluid component. Equation (A8) is correct in the two limiting regimes but is not very accurate in between, at least for the case of a single long circular cylindrical channel.

We reduce these eight simultaneous linear differential equations to eight algebraic equations in the usual way by assuming $exp(i\omega t)$ time dependences for the nine variables $I_s$, $I_n$, $P_1$, $P_2$, $T_1$, $T_2$, $x_1$, $x_2$, and $V$. For convenience, we introduce $\omega_1^2 \equiv k_1/m_1$, $\omega_2^2 \equiv k_2/m_2$, and $M_n \equiv L_n - (\rho/\rho_n)(iR_n/\omega)$. We regard the first eight variables as unknowns and $V$ as a known. In what follows, we use the equation numbers (A1) through (A8) to refer to the algebraic equations, rather than to the original differential equations. Zero subscripts denote the complex amplitudes of the variables.

At this point we could correct our treatment of the normal fluid equation of motion, at least for the case of a single long circular cylindrical channel, by following the more accurate treatment given in Ref. 37. The result would be the replacement of our $M_n$ by a more accurate complex function of $\omega$ having the same limiting forms as $M_n$, as $\omega \to 0$ and as $\omega \to \infty$. For our present purposes, we simply note that the absolute values of both our $M_n$ and its more accurate replacement are never less than $L_n$.[37]

The resonance frequencies of the system are determined by the secular equation, the equation that results from eliminating all eight independent variables from among the eight algebraic equations in the absence of $V_0$. On physical grounds we expect to find four resonances, resulting from the coupling of the two free-diaphragm resonances to the two Helmholtz-like resonances that the fluid would have in the absence of diaphragm flexibility and normal-fluid viscosity, one with comoving normal fluid and superfluid components, the other with countermoving components.[38]

Cumbersome as the set of equations is to handle, it is possible to develop the secular equation in the following manageable form:

$$\left[ \frac{\rho_s}{L_s}\left(1 - \frac{\alpha s T}{c}\right) + \frac{\rho_n}{M_n}\left(1 + \frac{\rho_s}{\rho_n}\frac{\alpha s T}{c}\right)\right] + \left[-\frac{\rho_s}{L_s}\left(1 - \frac{\alpha s T}{c}\right) + \frac{\rho_s}{M_n}\left(1 + \frac{\rho_s}{\rho_n}\frac{\alpha s T}{c}\right) + \frac{i\omega\rho\alpha}{scR_T}\right] \frac{\rho s (T_{20} - T_{10})}{P_{20} - P_{10}}$$

$$= \omega^2 \rho^3 \kappa \frac{\Omega_1\left(1 - \frac{\alpha^2 T}{\rho\kappa c} + \frac{\alpha_1}{1 - (\omega^2/\omega_1^2)}\right)\Omega_2\left(1 - \frac{\alpha^2 T}{\rho\kappa c} + \frac{\alpha_2}{1 - (\omega^2/\omega_2^2)}\right)}{\Omega_1\left(1 - \frac{\alpha^2 T}{\rho\kappa c} + \frac{\alpha_1}{1 - (\omega^2/\omega_1^2)}\right) + \Omega_2\left(1 - \frac{\alpha^2 T}{\rho\kappa c} + \frac{\alpha_2}{1 - (\omega^2/\omega_2^2)}\right)}, \quad (A9)$$

where

$$\frac{\rho s(T_{20} - T_{10})}{P_{20} - P_{10}} = \frac{\dfrac{\alpha s T}{c} - \dfrac{\rho \kappa s^2 T}{c}\dfrac{1}{\omega^2 \rho^3 \kappa \Omega_r}\left(\dfrac{\rho_s}{L_s} - \dfrac{\rho_s}{M_n}\right)}{1 - \dfrac{\rho_s}{\rho_n}\dfrac{\rho \kappa s^2 T}{c}\dfrac{1}{\omega^2 \rho^3 \kappa \Omega_r}\left(\dfrac{\rho_n}{L_s} + \dfrac{\rho_s}{M_n}\right) - \dfrac{i}{\omega\rho c \Omega_r R_T}}. \quad (A10)$$

Here we have introduced $\Omega_r \equiv (\Omega_1\Omega_2)/(\Omega_1 + \Omega_2)$, $\alpha_1 \equiv A_1^2/(\kappa k_1 \Omega_1)$, and $\alpha_2 \equiv A_2^2/(\kappa k_2 \Omega_2)$. In these expressions, $\kappa = (1/\rho)(\partial \rho/\partial P)_T$ is the isothermal compressibility, $\alpha = -(1/\rho)(\partial \rho/\partial T)_P$ is the isobaric thermal expansion coefficient, and $c = T(\partial s/\partial T)_P$ is the heat capacity per unit mass at constant $P$.

In order to obtain Eq. (A9), we have proceeded first by eliminating $x_{10}$ and $T_{10}$ among Eqs. (A1), (A3), and (A5), and $x_{20}$ and $T_{20}$ among Eqs. (A2), (A4), and (A6) with $V_0 = 0$, with the exception that the terms containing $T_{20} - T_{10}$ in Eqs. (A3) and (A4) are retained. The quantities $I_{s0}$, $I_{n0}$, and $T_{20} - T_{10}$ could then be eliminated between the results of these reductions to yield simple relations between $P_{10}$ and $P_{20}$ and thus between $P_{20} - P_{10}$ and $P_{10}$ and $P_{20}$ separately.

Using the relation between $P_{20} - P_{10}$ and $P_{10}$, together with Eqs. (A7) and (A8), the variables $P_{10}$, $I_{s0}$, and $I_{n0}$ have been eliminated in the result of the reduction of Eqs. (A1), (A3), and (A5) above to yield Eq. (A9).

In order to obtain Eq. (A10), we have gone back to Eqs. (A1) and (A3), once again eliminating $x_{10}$, but this time finding an expression for $I_{s0} - (\rho_s/\rho_n)I_{n0} + (T_{20} - T_{10})/(sTR_T)$ in terms of $P_{10}$ and $T_{10}$. From Eqs. (A2) and (A4), the variable $x_{20}$ has been eliminated to find an expression for $I_{s0} - (\rho_s/\rho_n)I_{n0} + (T_{20} - T_{10})/(sTR_T)$ in terms of $P_{20}$ and $T_{20}$. We then combined the two resulting equations to find $T_{20} - T_{10}$ in terms of $I_{s0}$, $I_{n0}$, and $P_{20} - P_{10}$. Using Eqs. (A7) and (A8) to eliminate $I_{s0}$ and $I_{n0}$, we then found Eq. (A10). The quantity $V_0$ never enters this derivation, so the result applies when $V_0 \neq 0$ as well as when $V_0 = 0$.

The two resonances of concern to us are, first of all, the one that evolves out of the rigid-diaphragm comoving-fluid mode, at an angular frequency we denote as $\omega_-$, and second, the one that evolves out of the free-diaphragm mode of the lower diaphragm, at an angular frequency we denote as $\omega_+$. For the mode at $\omega_-$, the terms in Eq. (A10) involving $1/(\omega^2 \rho^3 \kappa \Omega_r)$ times expressions containing $L_s$ and $M_n$ are of order unity, and for the mode at $\omega_+$, of order $< 10^{-1}$. The terms $\alpha s T/c$ and $\rho \kappa s^2 T/c$ are less than $10^{-4}$ at 1.0 K and remain less than $10^{-2}$ in magnitude up to 2.1 K. Hence the numerator of Eq. (A10) remains of order $10^{-2}$ or less under all conditions of interest. Thus if the denominator is of order unity or greater, $\rho s(T_{20} - T_{10})/(P_{20} - P_{10})$ will remain of order $10^{-2}$ or less throughout. The term $(\rho_s/\rho_n)(\rho \kappa s^2 T/c)$ in the denominator is less than $10^{-2}$ between 1.0 and 2.1 K, although it rises to 0.3 at 0.5 K. Thus the only possibility for the denominator to become less than of order unity would occur well below 1.0 K, where the real part of the second term in the denominator might possibly approach unity. However, in this region the normal fluid density is very small, normal fluid hydrodynamics is breaking down, and we assume that normal-fluid and thermal effects play little role.

If the $\rho s(T_{20} - T_{10})/(P_{20} - P_{10})$ term is neglected in Eq. (A9) and we temporarily treat the $\omega^2/\omega_2^2$ terms and $M_n$ as constants, the equation can be reduced in effect to a quadratic in $\omega^2$. Further, if the terms $\alpha s T/c$ and $\alpha^2 T/(\rho \kappa c)$ in Eq. (A9) are neglected, both of which have magnitudes less than $10^{-4}$ at 1.0 K and less than $10^{-2}$ up to 2.1 K, this quadratic yields the following two solutions, to first order in $\omega_{Heff'}^2/\omega_1^2$:

$$\omega_-^2 = \omega_{Heff'}^2 \left( 1 - \frac{\alpha_1}{1+\alpha_1} \frac{\Omega_{2eff}}{\Omega_{1eff} + \Omega_{2eff}} \frac{\omega_{Heff'}^2}{\omega_1^2} \right) \tag{A11}$$

and

$$\omega_+^2 = (1+\alpha_1)\omega_1^2\left(1 + \frac{\alpha_1}{1+\alpha_1}\frac{\Omega_{2eff}}{\Omega_{1eff}+\Omega_{2eff}}\frac{\omega_{Heff'}^2}{\omega_1^2}\right), \quad (A12)$$

where

$$\omega_{Heff'}^2 \equiv \frac{1}{\rho^3\kappa}\left(\frac{\rho_s}{L_s} + \frac{\rho_n}{M_n}\right)\frac{\Omega_{1eff}+\Omega_{2eff}}{\Omega_{1eff}\Omega_{2eff}}. \quad (A13)$$

Here $\Omega_{1eff} \equiv \Omega_1(1+\alpha_1)$ and $\Omega_{2eff} \equiv \Omega_2\left(1+\alpha_2/(1-\omega^2/\omega_2^2)\right)$. In each of Eqs. (A11) and (A12), $\Omega_{2eff}$ and $\omega_{Heff'}^2$ are to be evaluated at the resonant frequency in question. Thus these equations are implicit expressions for $\omega_-^2$ and $\omega_+^2$, respectively. In the low-temperature limit, $\omega_{Heff'}^2$ reduces to $\omega_{Heff}^2$ as defined by Eq. (7) in the main text, and thus Eqs. (A11) and (A12) reduce to Eqs. (6) and (12) in the main text.

It might be wondered where the other two resonances that were expected have gone. The highest-frequency resonance, whose angular frequency we denote as $\omega_{++}$, is eliminated in effect by the suppression of the $\omega^2/\omega_2^2$ terms. The lowest-frequency resonance, whose angular frequency we denote as $\omega_{--}$, is lost in dropping the $\rho s(T_{20}-T_{10})/(P_{20}-P_{10})$ term and its $\omega^2$ dependence. Although it is self-consistent to assume that this term is small for the resonances at $\omega_-$ and $\omega_+$, this term cannot be assumed to be small for the mode at $\omega_{--}$, in view of the fact that the denominator on the right-hand side of Eq. (A10) can become very small at low enough angular frequencies. The mode at $\omega_{--}$ was never observed.

In the derivation of Eqs. (A11) and (A12), the neglect of the $\rho s(T_{20}-T_{10})/(P_{20}-P_{10})$ term in Eq. (A9) might not be justified if the term $i\omega\rho\alpha/(scR_T)$ were to grow large relative to the first term in square brackets on the left-hand side. However, we estimate that under all of our conditions of interest, the ratio of the former to the latter is never larger than of order unity.

We turn now to the problem of determining the amplitudes of the unknown variables in response to some drive amplitude $V_0$. However, our principal interest is to determine the amplitude of the difference of the chemical potential per unit mass between chambers

$$\Delta\mu_0 = \mu_{20} - \mu_{10} = \frac{1}{\rho}(P_{20}-P_{10}) - s(T_{20}-T_{10}) \quad (A14)$$

in terms of $x_{10}$. We do not need to relate each unknown explicitly to $V_0$ as an intermediate step. We can derive the desired relationship from the original eight equations without approximation as follows:

$$\Delta\mu_0 = \frac{k_1}{\rho A_1}\left(1+\frac{\Omega_1}{\Omega_2}\frac{1-\frac{\alpha^2 T}{\rho\kappa c}+\frac{\alpha_1}{1-\left(\omega^2/\omega_1^2\right)}}{1-\frac{\alpha^2 T}{\rho\kappa c}+\frac{\alpha_2}{1-\left(\omega^2/\omega_2^2\right)}}\right)\left(1-\frac{\rho s(T_{20}-T_{10})}{P_{20}-P_{10}}\right)\left(1-\frac{\omega^2}{\omega_1^2}\right)x_{10}. \tag{A15}$$

As in the $V_0 = 0$ case, $\rho s(T_{20}-T_{10})/(P_{20}-P_{10})$ is given by Eq. (A10).

In order to derive Eq. (A15), we began, as in deriving Eq. (A9), by eliminating $x_{10}$ and $T_{10}$ but retaining $T_{20} - T_{10}$ among Eqs. (A1), (A3), and (A5). Using Eqs. (A7) and (A8), we then eliminated $I_{s0}$ and $I_{n0}$ from the result to obtain a relationship involving $P_{20} - P_{10}$, $T_{20} - T_{10}$, and $P_{10}$. This relationship was combined with Eqs. (A5) and (A14) to relate $\Delta\mu_0$ to $x_{10}$ in terms of $\rho s(T_{20}-T_{10})/(P_{20}-P_{10})$. Finally, this latter relation was simplified by use of Eq. (A9), although this last step restricts the applicability of the result to one of the resonant frequencies.

If, as in the case of the secular equation, we neglect the terms $\rho s(T_{20}-T_{10})/(P_{20}-P_{10})$ and $\alpha^2 T/(\rho\kappa c)$, we find

$$\Delta\mu_0 \cong \frac{k_1}{\rho A_1}\frac{\Omega_{1\mathit{eff}}+\Omega_{2\mathit{eff}}}{\Omega_{2\mathit{eff}}}\left(1-\frac{1}{1+\alpha_1}\left(1+\frac{\alpha_1\Omega_{2\mathit{eff}}}{\Omega_{1\mathit{eff}}+\Omega_{2\mathit{eff}}}\right)\frac{\omega^2}{\omega_1^2}\right)x_{10}, \tag{A16}$$

which appears as Eq. (15) in the main text.

For the case of the partially-filled upper chamber, we modify the eight original equations as follows. The quantity $P_2$ is assumed to equal zero throughout. Equations (A2) and (A4) become, respectively,

$$I_s + I_n = \left(\frac{\partial\rho}{\partial T}\right)_P W_2\frac{dT_2}{dt}+\rho\frac{dW_2}{dt}, \tag{A17}$$

$$\frac{\rho s}{\rho_n}I_n - \frac{T_2-T_1}{TR_T} = \left(\frac{\partial(\rho s)}{\partial T}\right)_P W_2\frac{dT_2}{dt}+\rho s\frac{dW_2}{dt}, \tag{A18}$$

where $W_2$ is the volume of fluid in the upper chamber. Equation (A6) drops out, leaving us with seven equations in the seven unknowns $I_s$, $I_n$, $P_1$, $T_1$, $T_2$, $x_1$, and $x_2$. It is unclear how best to introduce a drive term, the fluid motion at least to some extent being excited by the shaking of the entire cell. We will rely on the result from the filled-chamber case that the relation between $\Delta\mu_0$ and $x_{10}$ is independent of the drive and can be found satisfactorily without a drive term.

The resulting secular equation takes the same form as Eq. (A9) with $P_{20}=0$ in the limit that $\Omega_2 \to \infty$. In this case $\rho s(T_{20}-T_{10})/(-P_{10})$ is given by the same expression that Eq. (A10) gives for $\rho s(T_{20}-T_{10})/(P_{20}-P_{10})$, except that $\Omega_r$ is replaced by $\Omega_p \equiv (\Omega_1 W_2)/(\Omega_1+W_2)$. As a consequence, if we proceed as we did for the completely-filled-cell case and if we neglect $\rho s(T_{20}-T_{10})/(-P_{10})$ in the secular equation as well as the same other small terms, we obtain expressions for $\omega_-^2$ and $\omega_+^2$ that are of the same form as Eqs. (A11), (A12), and (A13) in the limit that $\Omega_{2eff} \to \infty$. In this partially-filled-cell case, we designate the counterpart of $\omega_{Heff'}^2$ as $\omega_{Heff*'}^2$. The resulting relation between $\Delta\mu_0$ and $x_{10}$ at the resonances takes the same form as Eq. (A15) in the limit that $\Omega_2 \to \infty$ and with $P_{20}=0$. Employing the same approximations used in obtaining Eq. (A16) from (A15), we obtain the simpler expression

$$\Delta\mu_0 \cong \frac{k_1}{\rho A_1}\left(1-\frac{\omega^2}{\omega_1^2}\right)x_{10}, \tag{A19}$$

which appears as Eq. (16) in the main text.